\documentstyle [12pt]{article}
\newcommand{\doublespace}{
    \renewcommand{\baselinestretch}{1.6}\large\normalsize}

\def\agl#1{\begin{equation} \label{#1}}
\def\egl{\end{equation}}

\def\pr{^{\>\prime}}
\def\bra#1{\big\langle #1 \big\vert}
\def\ket#1{\big\vert #1 \big\rangle}
\def\d3{d^3}
\def\inv#1{\frac{1}{#1}}

\def\t{\theta}
\def\f{f(\t)}

\def\rp{r \pr}
\def\gl{e^{i\chi(b)}}

\begin{document}

\begin{titlepage}

\pagestyle{empty}
\begin{flushright}
{\small August, 1996}
\end{flushright}
\vspace{0.5cm}
\doublespace

\begin{center}
\begin{Large}
{\bf{Effects of Strong Interaction on the Electromagnetic Dissociation}}
\end{Large}
\vskip 0.2in
A. M\"UNDEL and G. BAUR\\
{\small{\it IKP (Theorie), Forschungszentrum J\"ulich, D-52425 J\"ulich, Fed. Rep. Germany.}}
\end{center}

\parindent=0pt
\frenchspacing

\begin{abstract}

We investigate effects of strong interactions on the electromagnetic
dissociation of nuclei in heavy ion collisions.

We start from the eikonal approach to the equivalent photon method to
describe the electromagnetic contributions to the cross section of
peripheral collisions. We summarize some results of this approach and
characterize recent experiments in "universal plots".

In the second part of this work, we give a straight forward method to
include transitions induced by the strong interactions between the
ions. We introduce different methods to obtain the nuclear
transition potential, and study the behavior of the resulting nuclear
contributions.
 
\end{abstract}
\end{titlepage}
 

\newpage
\pagestyle{plain}
\baselineskip 16pt
\vskip 48pt
\doublespace

\section{Introduction}

The Coulomb field of a fast moving heavy ion is an intense source of
quasi real photons \cite{{Weiz},{Wil},{PR163}}. With raising
projectile energy this photon spectrum becomes harder and there exist
a variety of applications in nuclear-- as well as astrophysics, like
the excitation of giant resonances or the Coulomb dissociation of the
ions \cite{Baur1}.

The advantage of using heavy ions to study this processes are the
large cross section, which make the Coulomb dissociation an
interesting tool to study even multi--phonon resonances
\cite{{Bert3},{Emling}} or the astrophysical S--factor for reactions
like $^7Be(p,\gamma)^8B$ \cite{Baur1}. In the latter case phase-space
considerations and the large number of equivalent photons highly
favor the Coulomb dissociation compared to the direct measurement of
the radiative capture reaction
\cite{{Baur2},{Esbensen}}.

A powerful but simple tool to describe the Coulomb dissociation of
heavy ions is the equivalent photon method \cite{PR163}. On the other
hand this simple approach does not take into account excitations due
to the strong interaction. Therefore it is interesting and necessary
to study these effects. Using the hydro-dynamical model of Bohr and
Tassie for the nuclear transition densities or potentials
\cite{Tassie}, one can connect the nuclear interaction to the optical
potential and the deformation parameter and therefore to the
electromagnetic transition matrixelements. This approach allows us to
study these contributions in various cases of physical interest.

\section{Equivalent photon spectrum \newline in a Glauber approach}

In the following we consider a situation where a projectile (particle
1) with charge $Z_1$ excites a target (particle 2) with charge $Z_2$
from a state $\ket{I_i=0,M_i=0}$ to the excited state $\ket{I,M}$ in a
peripheral collision via the exchange of one quasi real photon of
energy $E_{\gamma} = \hbar \omega$. The velocity $\vec \beta =
\frac{\vec v}{c}$ of the projectile is in z direction and we use
$\gamma = \left( 1 - \beta ^2\right)^{-\frac{1}{2}}$. The scattering
angle of the projectile is denoted by $\theta$.  The inelastic
scattering amplitude in the eikonal approximation $\f$
is than given by \cite{{Glauber},{Bert1}}:
\begin{equation} \label{f}
\f =\frac{i k}{2 \pi \hbar v} \int \d3 r \ \d3 \rp \ 
e^{i\vec q \cdot \vec r} \ \gl  
\bra{\Phi_f(\vec \rp)} V_{int}(\vec r , \vec \rp) \ket{\Phi_i(\vec \rp)},
\end{equation}
where $\vec q$ denotes the momentum transfer and $k$ is the wavenumber
of the incoming projectile. We define $\vec r$ to be the separation
of the centers of mass of the two nuclei, and $\vec
\rp$ to be the intrinsic coordinate of the target nucleus.

The Glauber phase $\chi (b)$ is the sum of a Coulomb and a nuclear
part. If we assume the 'sharp cut--off' model for the nuclear part
\begin{equation} \label{sharp} 
e^{i\chi_N(b)} = 
\cases{
0& if $b \leq R = R_1 + R_2$ \cr
1& if $b > R$},
\end{equation}
where $R_i = 1.2 \ fm \ A^{1/3}$, ($i=1,2$) is the radius of the i'th
nucleus, the Coulomb phase is given by
\begin{equation} \label{chic}
\chi_C(b) = 2 \ \eta \ ln(k b). 
\end{equation}
Here $\eta = \frac{Z_1Z_2 e^2}{\hbar v}$ is the Sommerfeld parameter.

Using standard methods \cite{PR163} and defining the reduced
electromagnetic transition probabilities $B(\pi l)=B(\pi l,I_i
\rightarrow I_f)$ \cite{AW1} we can write the cross section in the
 form
\begin{equation} \label{cross}
\frac{d \sigma}{d \Omega} 
= \sum_{\pi l}\int \frac{d E_{\gamma}}{E_{\gamma}} \ 
\sigma_{E_{\gamma}}^{\pi l}(E_{\gamma})
\ \frac{d n^{\pi l}}{d\Omega}. 
\end{equation}
In the above formula $\sigma_{\gamma}^{\pi l}(E_{\gamma})$ is the
 photo-nuclear absorption cross section for a given multipolarity
\cite{PR163}
\begin{equation} \label{photo}
\sigma_{\gamma}^{\pi l}(E_{\gamma}) = 
\frac{(2 \pi)^3(l+1)}{l[(2l+1)!!]^2}
\rho_F(E_{\gamma}) \ (\frac{\omega}{c})^{2l-1} B(\pi l),
\end{equation}
where $ \rho_F(\omega)$ is the final state density.

Under these assumptions the equivalent photon number per unit solid
angle for a given multipolarity $\frac{d n^{\pi l}}{d\Omega}$ is given
by:
\begin{equation} \label{nc} 
\frac{d n^{\pi
l}_C}{d\Omega} = Z_1^2 \alpha \left(\frac{\omega k}{\gamma v}\right)^2
\frac{l[(2l+1)!!]^2}{(2 \pi)^3(l+1)} 
\sum_m  \left\vert G_{\pi l m}
\left(\inv{\beta}\right) \right\vert ^2 \big\vert \Omega_m^C(q_T)
\big\vert^2,  
\end{equation}
Here the transverse momentum transfer $q_T\approx k \theta$ is the only quantity,
which depends on the scattering angle. Defining the variables
\begin{eqnarray}\label{var}
x=\frac{b}{R} &;& \xi=\frac{\omega R}{\gamma v}; \nonumber \\
\Theta_{diff}=\inv{kR} &;&\theta_{gr} = 2 \eta\theta_{diff},
\end{eqnarray} 
$\Omega_m^C$ can be written as:
\begin{equation} \label{omega2}
\Omega_m^C(\theta) = R^2 \int_1^\infty x\ dx
J_m(\frac{\theta x}{\theta_{diff}})
K_m(\xi x) \ e^{2i\eta ln (\frac{x}{\theta_{diff}} )}. 
\end{equation}

In the case of the classic trajectory of the projectile ($\eta
\rightarrow \infty$),  one can use the saddle 
points method to solve the integral in eq. (\ref{omega2})\cite{Bert1}:
\begin{equation} \label{omegasc}
\Omega_m^{sc}(\theta) = R^2 \frac{y^2}{2\eta} 
\ e^{i\tilde \phi} \ K_m(\xi y) \ ,
\end{equation}
where the variables $y= \frac{\theta_{gr}}{\theta}$ and $\tilde
\phi = \frac{1}{2} \pi (m + 1) + 2 \eta \left(
ln\left(\frac{2\eta}{\theta}\right) - 1 \right)$ where introduced.

Expression (eq. \ref{omega2}) for $\Omega_m$ mainly depends on two
parameters, the interaction strength $\eta$ and the adiabaticity
parameter $\xi$, which is the ratio of impact time to transition time
in a grazing collision.

It is interesting to compare $\vert 2 \eta \frac{\Omega_m}{R^2}\vert
^2$ as a function of $\frac{\theta}{\theta_{gr}}$ for fixed $\xi$ but
different $\eta$ as shown in the 'universal plots' in
figs. (1) and (2). The thick solid line shows the semi
classical limit according to eq. (\ref{omegasc}) while the dashed and
dotted lines are the solutions of eq (\ref{omega2}) for different values
of $\eta$. At $\frac{\theta}{\theta_{gr}}=1$ the semi classical limit
drops to zero while the full solutions extend to higher angles. At
small angles the full solution shows diffraction effects, but for a
large interval of angles the full solution for sufficiently large
$\eta$ is a fast oscillating function around the semi classical limit
(dotted line). Only for small $\eta$ (dashed line) the full solution
differs substantially from the semi classical limit.

For the above reason the semi classical solution for the inelastic
scattering amplitude (eq. \ref{f}) provides in many relevant cases
already a good description of the experimental
situation. Fig. (3) classifies some existing experiments in a
$\xi$--$\eta$--plot where the experiments well described by the semi
classical solution are at large $\eta$ and small $\xi$.

To complete the discussion of the sharp cut-off model we want to give
the expression for the total cross section, which is obtained by
integrating eq. (\ref{cross}) over the solid angle and using the
closure relation for the Bessel functions. One finds:
\begin{equation}\label{total}
\sigma_{tot} = 2 \pi \left(\frac{Ze}{\hbar c}\right)^2 
\left(\frac{\omega}{\gamma v}\right)^2
\sum_{\pi l m}\left(\frac{\omega}{c}\right)^{2(l-1)}
|G_{\pi l m}(\frac{1}{\beta})|^2 B(\pi l)
R^2 \int_{1}^{\infty} x \ dx \ K_m^2(\xi x).
\end{equation}
Notice, that this expression is independent of $\eta$.

\section{The nuclear excitation part}

The sharp cut--off model, as it was introduced in the last section,
takes nuclear effects into account only in a very simplified way.  A
more realistic model for the nuclear contributions will influences the
inelastic scattering amplitude (eq. \ref{f}) in two different
ways. First it leads to an additional transition potential in
eq. (\ref{f}) of the form $\bra{\Phi_f(\vec \rp)}V_{N}\ket{\Phi_i(\vec
\rp)}$, where $V_N$ is the nuclear potential between the two
nuclei. In addition it will change the nuclear phase due to
\begin{equation} \label{glauber}
\chi_N(b) = 
- \inv{\hbar v} \int_{-\infty}^{\infty} V_{N}(\sqrt{b^2 + z^2}) dz ,
\end{equation}
which, in contrast to the sharp cut--off model, now has a finite real
and imaginary part.  This leads to a "smoother" cut--off than in the
sharp cut--off model and a modification of the Coulomb potential $V_C$
due to the penetration of the two charge distributions. In the following
we will neglect the effect on the Coulomb potential but keep
eq. (\ref{glauber}), and concentrate on the nuclear induced
transitions.

There exist different approaches to obtain the nuclear transition
potential (eq. \ref{f}) $\bra{\Phi_f(\vec \rp)}V_{N}\ket{\Phi_i(\vec
\rp)}$. In this work we assume a collective nuclear model. 

First we want to describe the approach via the transition
densities. In this case the nuclear transition potential can be
calculated using the folding formalism \cite{Bert2}:
\begin{equation}\label{matn}
\bra{\Phi_f(\vec \rp)}V_N(\vec r,\vec \rp)\ket{\Phi_i(\vec \rp)} 
= \left\langle t_{NN}(E) \right\rangle \ \rho_1(\vec r - \vec \rp) 
\ \left\langle I,M \vert \rho_2(\vec \rp) \vert 0,0 \right\rangle.
\end{equation}
Here $\rho_i, \ (i=1,2)$ denotes the charge density of the i'th
nucleus. For the energy dependent $t_{NN}$--matrix see
ref. \cite{Bert2}.  To determine the transition density matrix
elements $\left\langle I,M
\vert \rho_2(\vec \rp) \vert 0,0 \right\rangle$ we assume the
hydro-dynamical model of Bohr and Tassie
\cite{Tassie}:
\begin{equation} \label{BTM}
\bra{I,M}\rho_2(\vec \rp)\ket{0,0} = 
\frac{\beta_I R_2^{2-I}}{\sqrt{2I+1}} {\rp}^{(I-1)} 
\frac{d\rho_2(\vec \rp)}{d\rp} \ Y^*_{IM}(\hat \rp).
\end{equation}
The deformation parameter $\beta_l$ is related to the electromagnetic
transition matrixelements $M(E l,m)$ by:
\begin{equation} \label{elmagmat}
M(El,m) = \sqrt{2l+1} \beta_l R^{2-l}_2 
\int _0^{\infty}dr \ r^{2l} \rho_2(r)\ \delta_{M,m}
\end{equation}
Under these assumptions the inelastic scattering amplitude
(eq. \ref{f}) can be written as:
\begin{eqnarray} \label{fg}
\f = &\sum_{\pi l m}& \frac{i^{m+1} k}{\hbar v} 
\ e^{-im\alpha} \ \bra{I,M}M(\pi l,m)\ket{0,0} \nonumber \\
&\int_{0}^{\infty}& b \ db\ e^{i(\chi_C(b) +\chi_N(b))}
J_m(q_T b) \ 
\left\lbrace \Gamma_C^{\pi l m}(b) 
+ \Gamma_N^{\pi l m}(b) \right\rbrace,
\end{eqnarray}
where $\alpha$ is the azimuthal scattering angle.
Since we neglect the corrections of the Coulomb part due to the
penetration of the two particles, $\Gamma^{\pi lm}_C$ is still assumed
to be given by: 
\begin{equation} \label{gammac}
\Gamma^{\pi lm}_C (b)= \frac{Z_1 e}{\gamma} \ (\frac{\omega}{c})^l\ \sqrt{2l+1}
G_{\pi lm} \left(\inv{\beta}\right) 
\ K_m\left(\frac{\omega b}{\gamma v}\right) .
\end{equation}

For the sake of mathematical simplicity we consider a Gaussian shaped
charge distribution for the two nuclei \cite{Bert2} to calculate
$\Gamma^{\pi lm}_N$:
\begin{equation} \label{rho}
\rho_i(r) = \rho_i(0) e^{-\frac{r^2}{R_{G_i}^2}}
\ ; \ \ 
\rho_i(0) = 0.085 fm^{-3} e^{\frac{R_i}{2a_i}}.
\end{equation}
We have used the notation $R_{G_i}^2 = 2a_iR_i$, where $a_i$ is the
diffuseness parameter of the i'th nucleus. This choice for the charge
distribution allows us to perform most of the calculations in an
analytical way.

Under the above assumptions the nuclear phase (eq. \ref{glauber}) is
given by \cite{ Karol}
\begin{equation} \label{chin}
\chi_N(b) = \left\langle t_{NN}(E) \right\rangle 
\ \rho_1(0) \ \rho_2(0) \left\lbrace \frac{R_{G_1}^3 
R_{G_2}^3}{R_G^2} \right\rbrace e^{-b^2/R_G^2} ,
\end{equation}
where $R_G^2 = R_{G_1}^2 + R_{G_2}^2$. To take into account the
singularity $\chi_C(b=0)$ (eq. \ref{chic}), we modify the
Coulomb phase \cite{Bert1}:
\begin{equation}\label{chinmod}
\chi_C(b) = 2\eta\left\lbrace ln(kb) 
+ \frac{1}{2}E_1\left(\frac{b^2}{R^2_G}\right) \right\rbrace.
\end{equation}
In the above formula $E_1$ denotes the exponential integral. 

Calculating the nuclear transition density as given in
eq. (\ref{BTM}) and using eq. (\ref{elmagmat}) we find for
$\Gamma_N^{\pi lm}(b)$ :
\begin{eqnarray}\label{gammagauss}
\Gamma^{E lm}_N (b)&=& \pi \ \frac{\rho_1(0)}{e}  \ 
\frac{2^{l+2}}{(2l+1)!!} \ \frac{R_{G_1}^3}{R_G^{2l+3}} \left\langle
t_{NN}(E) \right\rangle \nonumber \\
&&\int dz\ r^l \ e^{i
\frac{\omega}{v}z} \ e^{-\frac{r^2}{R_G^2}} \ Y_{lm}(\theta_{\hat r},0) \delta_{I,l},
\nonumber \\
\Gamma^{M lm}_N (b)&=& 0. 
\end{eqnarray}
 
The Gauss parameterization for the charge densities is a relatively
good approximation for light nuclei, but fails for heavy
nuclei. Starting from more realistic densities immediately makes it
necessary to solve the folding integral (eq. \ref{matn})
numerically. To avoid these difficulties one can start from optical
potential rather then their corresponding densities. This formalism is
described in ref. \cite{Satchler1}.

For deformed optical potentials of the form $V_N(R(\theta,\phi),r)$,
where $R(\theta,\phi) = R_0(1 + \sum_{\lambda\mu}
a_{\lambda\mu}Y^*_{\lambda\mu}(\theta,\phi))$, that depend only on the
distance $(R(\theta,\phi)-r)$, the nuclear transition potential
$\bra{I,M}V_N\ket{0,0}$ is of the form \cite{SatchlerII}
\begin{equation}\label{transpot}
\bra{I,M} V_N \ket{0,0} = \frac{-1}{\sqrt{2I+1}}\beta_I \ R_2
\frac{dV_N(R(\theta,\phi)-r)}{dr} Y^*_{IM}(\theta,\phi).
\end{equation}
In both, the Tassie model and the Bohr Mottelson model
\cite{BM}, $\beta_I$ is than given by
\begin{equation}\label{beta}
\beta_I = \frac{4\pi}{3}\frac{1}{Z_2R^I_2}\sqrt{\frac{B(EI)}{e^2}}.
\end{equation}
Using eqn. (\ref{transpot}, \ref{beta}, \ref{fg}, and \ref{f}) we find for 
$\Gamma_N^{Elm}(b)$ :
\begin{equation}\label{gammadv}
\Gamma^{E lm}_N (b)= \frac{-4\pi}{3Z_2eR^{l-1}_2}
\int dz\ \ e^{i
\frac{\omega}{v}z} \ \frac{dV_N(R(\theta,\phi)-r)}{dr} \ Y_{lm}(\theta_{\hat r},0) \delta_{I,l}, 
\end{equation}
with $r^2=b^2 + z^2$.

This result can be compared to the Distorted Wave Born Approximation
(DWBA), see for example ref. \cite{Blair}. In this formalism the
nuclear part of the inelastic scattering amplitude for a transition
from a state $\ket{I_i=0,M_i=0}$ to the exited state $\ket{I,M}$ is given by
\begin{eqnarray}\label{DWBA}
f^{DWBA}_{N,I,M}(\theta) &=& \frac{-2ik^2}{\hbar v} \sum_\lambda \sum_{\lambda',\mu'}
\sqrt{4\pi} \ i^{\lambda - \lambda'} \ e^{i(\sigma_\lambda +
\sigma_{\lambda'})}
\sqrt{2\lambda'+1} \ \frac{R_2 \beta_I}{\sqrt{2I+1}} \ \frac{1}{kk'} R^I_{\lambda \lambda'} 
\nonumber \\ && \sqrt{\frac{2I+1}{4\pi}}
(\lambda' \ \mu' \ I \ M | \lambda \ 0)(\lambda' \ 0 \ I \ 0 | \lambda
\ 0) \ Y_{\lambda' \mu'}(\theta,0) ,
\end{eqnarray}
where $k$ ($k'$) is the wave number of the incoming (outgoing) wave. The
radial matrixelement $R^I_{\lambda' \lambda}$ has the form
\cite{Blair}:
\begin{equation}\label{DWBAradmat}
R^I_{\lambda \lambda'} = \int
d^3r f_{\lambda'}(k'r) \frac{dV_N(R(\theta,\phi)-r)}{dR_0} f_{\lambda}(kr).
\end{equation} 
For small scattering angles $\theta$ and large values for the angular
momentum of the partial waves $\lambda,\lambda'$ one can show that eq. (\ref{DWBA})
is equivalent to the derived Glauber approach
(eqn. \ref{fg}, \ref{gammadv}).

This connection of the Glauber model to the DWBA can be seen as
following.  In eq. (\ref{DWBAradmat}) the radial Coulomb wave function
$f_\lambda$ in the WKB approximation is given by
\cite{{Blair},{AlderII}}
\begin{equation}
f_{\lambda} \approx e^{i\delta_{\lambda}} (k^2/g(r))^\frac{1}{4} sin\Phi 
\end{equation}
where
\begin{eqnarray}\label{Coulombwf}
\Phi &=& \frac{\pi}{4}+\int_{r_0}^r \sqrt{g(r)}\ dr
\nonumber \\
g(r) &=& k^2-\frac{2\eta k}{r}-\frac{\lambda(\lambda+1)}{r^2}
\approx k^2-\frac{\lambda(\lambda+1)}{r^2} 
\end{eqnarray}
and $r_0$ is defined by $g(r_0) = 0$.

Assuming $R^I_{\lambda' \lambda} \approx R^I_{\bar \lambda \bar
\lambda}$ with $\bar \lambda = \frac{1}{2} (\lambda + \lambda')$ for
sufficiently large values of $\lambda$ one can follow the calculations
from ref. \cite{AlderII}, neglecting the rapidly scillating terms, and
obtain
\begin{equation} \label{radmat}
R^I_{\bar \lambda \bar \lambda} = e^{i\chi_N(b)} \frac{1}{4}
\int_{-\infty}^\infty dz \ e^{i\frac{\omega}{v}z} \frac{dV_N(R(\theta,\phi),b,z)}{dR_0},
\end{equation}
where we have used $e^{2i\delta_{\bar \lambda}} = e^{i\chi_N(b)}$ and
neglected the Coulomb part in eq. (\ref{Coulombwf}), which leads to straight
line trajectories instead of Rutherford trajectories. The $\bar
\lambda$--dependence of $R^I_{\bar \lambda \bar \lambda}$ is contained
in the impact parameter dependence of eq. (\ref{radmat}), since one
has $\bar \lambda + \frac{1}{2} \approx kb$. Making use of the
approximation
\begin{equation}
Y_{\lambda' -M}(\theta,0) \approx \sqrt{\frac{2\lambda' +1}{4\pi}} J_M(kb\theta),
\end{equation}
in the limit of $\theta \ll 1$ and replace $\sum_\lambda \rightarrow
k\int db$ on has
\begin{eqnarray}
f^{DWBA}_{N,I,M} (\theta) &=& \frac{-ik}{\hbar v} 
\frac{2}{\sqrt{\pi}} R_2 \beta_I \int b\ db 
\ e^{i\chi_N(b) + \chi_C(b)} J_M(kb\theta) \nonumber \\
&& \sum_{\lambda'} i^{\lambda - \lambda'} 
(\lambda' \ -M \ I \ M | \lambda \ 0)(\lambda' \ 0 \ I \ 0 | \lambda
\ 0) \frac{1}{4} \int_{-\infty}^\infty dz 
\ e^{i\frac{\omega}{v}z} \frac{d V_N(R(\theta,\phi),b,z)}{d R_0}.
\end{eqnarray}
Here we have also identified $e^{i(\sigma_\lambda + \sigma_{\lambda'})} 
\approx e^{2i\sigma_{\bar \lambda}} = e^{i\chi_C(b)}$. Finally we make 
use of the approximation \cite{Blair}
\begin{equation}
\sum_{\lambda'} i^{\lambda-\lambda'} 
(\lambda' \ -M \ I \ M | \lambda \ 0)(\lambda' \ 0 \ I \ 0 | \lambda
\ 0) \approx i^{-M} \sqrt{\frac{4\pi}{2l+1}} 
Y_{IM}(\frac{\pi}{2},0),
\end{equation}
for $(I+M)$ even and zero otherwise, and obtain for optical potentials
$V_N(R(\theta,\phi),r)$ that depend only on $(R(\theta,\phi)-r)$
\begin{equation}
f^{DWBA}_{N,I,M} (\theta) = (-1)^M \frac{i^{M+1}k}{\hbar v} 
\frac{R_2 \beta_I}{\sqrt{2I+1}} 
\int b \ db e^{i\chi(b)} J_M(kb\theta) 
\int _{-\infty}^\infty dz \ e^{i\frac{\omega}{v}z}
\frac{dV_N(R(\theta,\phi),b,z)}{dr} Y_{IM}(\frac{\pi}{2},0).
\end{equation}
This result is identical to eqn. (\ref{fg}, \ref{gammadv}) for a
surface peaked nuclear interaction ($Y_{IM}(\theta_{\hat r},0) \approx
Y_{IM}(\frac{\pi}{2},0)$) apart from an overall phase, which doesn't
influence the cross section.

We want to stress, that we made essential use of the assumption, that
the angular momentum of the contributing partial waves $\lambda$ and
$\lambda'$ are large, and that the cross section is limited to small
scattering angles, which both are sufficiently fulfilled for most
heavy ion collisions.

To apply eqn. (\ref{fg}, \ref{gammadv}) for the nuclear contribution
we want to assume a Woods--Saxon parameterization for the
nuclear potential $V_N = U(r) + iW(r)$ where
\begin{equation}\label{wspot}
U(r) = -U_0 \left(e^{\frac{r-R_U}{a_U}}+1\right)^{-1}
\ ; \ \
W(r) = -W_0 \left(e^{\frac{r-R_W}{a_W}}+1\right)^{-1}.
\end{equation}

In fig (4) we show the results with the above formalism for the
Coulomb dissociation of $^8B$. This reaction has recently been studied
experimentally \cite{Moto2} in order to determine the astrophysical
S--factor of the $^7B(p,\gamma)^8B$ reaction \cite{{Moto1},{Moto2}}. The figure
shows the cross section for a $51.9 MeV/nucleon$ $^8B$ beam on a
$^{208}Pb$ target. The excitation energy is assumed to be $1.2
MeV$. The solid line shows the total cross section ($E1 + E2$), the
dashed line the sum of the Coulomb and nuclear contribution for $l=2$
($E2_{C+N}$), and the dotted line gives the $E2_N$ contribution
only. We used $S_{17}(E1)=20 eV b$ and $S_{17}(E2)=10 meV b$. For the
Woods--Saxon potential we assumed the following parameters \cite{Barrette},
\begin{eqnarray}\label{wspar}
U_0 = 50 MeV &;& W_0 = 57 MeV\ ;\nonumber \\
R_U=R_W=8.5 fm &;& a_U=a_W=0.8 fm.
\end{eqnarray}
One should bare in mind that the use of the collective model for this
$^8B \rightarrow ^7Be + p$ continuum transition is not well
justified. A diffraction dissociation approach would certainly be more
appropriate, see e.g. ref. \cite{Kai}.

A more schematic model for the optical potential $V_N=U_N + i W_N$ is
given by the "ramp" potential,
\begin{equation}\label{rampe}
U(r)=
\cases{
-U_0 & if $r < R_U^-$ \cr
-\frac{U_0}{\Delta_U} (R_U^+-r) 
      & if $ R_U^- \leq r \leq  R_U^+$ \cr
0 & if $r > R_U^+$},
\end{equation}
where $R_U^{\pm} = R_U \pm \frac{\Delta_U}{2}$. The imaginary part
$W_N$ is given in a similar way.

Fig. (5) compares the nuclear phase $|e^{i\chi_N(b)}|^2$ and
$|\Gamma^{lm}_N(b)|^2$ for Woods--Saxon (solid line) and the "ramp"
potential (dashed line). The functions where calculated at $\theta =
2.0^{o}$, for $l=2$, $m=0$, and are plotted as a function of
$b/R$. For the Woods--Saxon potential we used $U_0 = 50 MeV$, $W_0 =
57 MeV$, $R_U=R_W=8.5 fm$, and $a_U=a_W=0.8$ and the same parameters
for the "ramp" potential but $\Delta_U=\Delta_W=4 fm$. One finds, that
for an adequate choice of parameters both models give similar results.

Because the product $|e^{i\chi_N(b)}\Gamma^{lm}_N(b)|$ is a good
measure for the nuclear transition strength, we want to deduce some
basic features of this quantity.  The magnitude of the real part $U_0$ is
directly proportional to the magnitude of $|\Gamma^{lm}_N(b)|$, while
the imaginary part $W_N$ mainly influences the shape of
$|e^{i\chi_N(b)}|^2$. Namely for decreasing $W_0$ the contributions
for $b<R-\Delta_W$ in eq. (\ref{fg}) becomes important. This leads to
an increasing nuclear contribution, while the oscillations of the
Coulomb contribution around the semi classical limit
(figs. 1 and 2) decrease.  For the radii a choice
like $R_U > R_W$ will lead to a larger overlap in fig. (5)
and therefore to a larger value for
$|e^{i\chi_N(b)}\Gamma^{lm}_N(b)|$.

This shows, that the magnitude of the nuclear contribution can vary
over a wide range if one uses
different sets of parameters for the optical potential.

For sufficiently large $W_0$ and small $\Delta$ one can deduce a very
simple expression for the "ramp" model. For simplicity we assume
$R_U=R_W=R$ and $\Delta_U=\Delta_W=\Delta$. The nuclear phase
is then approximately given by
\begin{equation}\label{glramp}
e^{i\chi_N(b)} =
\cases{
0 & if $b < R_U^-$ \cr
\frac{b-R^-}{\Delta}
      & if $ R_U^- \leq b \leq  R_U^+$  \cr
1 & if $b > R_U^+$},
\end{equation}
Using this approximations, we find for $e^{i\chi(b)}
\Gamma^{Elm}_N(b)$:
\begin{eqnarray}\label{strenght}
e^{i\chi_N(b)} \Gamma^{Elm}_N(b) &=& 
\frac{4\pi}{3Z_2eR^{l+1}_2} \frac{b-R^-}{\Delta^2} (U_0 + iW_0) 
\nonumber \\
&&  Y_{lm}(\frac{\pi}{2},0) \frac{2v}{\omega}
sin\left(\frac{\omega z_{max}}{v}\right)\ \delta_{I,l},
\end{eqnarray}
for $ R_U^- \leq b \leq R_U^+$ and zero otherwise.  In
eq. (\ref{strenght}) we used $Y_{lm}(\theta_{\hat r},0) \approx
Y_{lm}(\frac{\pi}{2},0)$ and we have defined $z_{max} = \sqrt{{R^+}^2
- b^2}$. 

Expanding the sinus for small arguments and introducing the variables
$D=(R_1 - R_2)/R$, $x=b/R$, and $\delta = \Delta/R$ one can write
eq. (\ref{strenght})
\begin{eqnarray}\label{strenght2}
e^{i\chi_N(x)} \Gamma^{Elm}_N(x)  &=& 
\frac{8\pi}{3 Z_2e} \frac{2^l}{R^{l-2}} 
\frac{U_0 + iW_0}{(1-D)^{l+1}} \frac{1}{\delta}\ Y_{l,m}(\frac{\pi}{2},0)\nonumber \\
&& (x-1+\delta) \sqrt{(1+\frac{\delta}{2})^2-x^2}\ \delta_{I,l}.
\end{eqnarray}

To see the influence of the parameters $\delta$ and $D$
Fig. (6) shows $|f^{E22}_N(\theta)|^2$ obtained from
eq. (\ref{strenght})as a function of $\theta /\theta _{gr}$ for
different values of $\delta$ (solid line $\delta=0.5$, dashed line
$\delta=0.1$ and dotted line $\delta=0.02$). We used $D=0.5$,
$\xi=0.1$ and $\eta=9$. Fig (7) shows the same for fixed
$\delta=0.5$ but varying $D$ (solid line $D=0.5$, dashed line $D=0.3$
and dotted line $D=0$).

We want to stress, that $\Gamma^{Elm}_N(b)$ is nearly independent from
the excitation energy $\omega$, while $\Gamma^{Elm}_C(b)$ decreases with
increasing excitation energy due to the behavior of
$K_m\left(\frac{\omega b}{\gamma v}\right)$ (eq. \ref{gammac}).

Finally we want to discuss the limit of the sharp cut--off model for
the nucleus in the adiabatic limit ($k'=k$). In this limit
eq. (\ref{radmat}) becomes
\begin{equation}
R^I_{\bar \lambda \bar \lambda}(k'=k) = e^{i\chi_N(b)} \frac{1}{4}
\int_{-\infty}^\infty dz \ \frac{dV_N(R(\theta,\phi),b,z)}{dR_0},
\end{equation}
and one can derive a relation between this radial matrixelement and
the nuclear phase $e^{i\chi_N(b)}$
\begin{equation}
R^I_{\bar \lambda \bar \lambda}(k'=k) = \frac{i \hbar v}{4} 
\frac{d e^{i\chi_N(b)}}{d R_0},
\end{equation}
where $e^{i\chi(b)}$ is like the optical potential $V_N(R(\theta,\phi),r)$ a
function of the distance $(R(\theta,\phi)-r)$.  Using eq. (\ref{sharp}) for the
nuclear phase, one finds for the sharp cut--off model
\begin{equation}\label{nsharp}
f_{N,I,M}(\theta) = i^M \ e^{-iM\alpha} \ \beta_I \ k\ R_2\  R 
\ e^{i\chi_C(R)} \ Y_{LM}(\frac{\pi}{2},0) \ J_M(q_T R).
\end{equation}
Using the asymptotic expansion for the Bessel function for large
$q_TR$ one finds the typical oscillatory behavior for the nuclear
contribution as predicted by Blair \cite{Blair}.

Fig. 8 compares the results for the E2 nuclear cross section for the
different models for the same experiment as in Fig. 4. ($51.9
MeV/nucleon$ $^8B$ beam on a $^{208}Pb$ target). The solid line shows
the result according to eq. (\ref{gammadv}) for a Woods--Saxon optical
potential with the parameters as given in (\ref{wspar}). The dashed
line shows the result for the ramp model with $U_0=50MeV$,
$W_0=57MeV$; $R_U=R_W=8.5 fm$ and $\Delta_U=\Delta_W=4fm$ and the
dotted line shows the result for the sharp cut--off model
(eq. \ref{nsharp}). The sharp cut--off result shows the typical
Frauenhofer diffraction while both other models show a smoother
nuclear contribution and a phase shift at higher scattering angles as
compared to the sharp cut--off model due to the diffuse edge of the
optical potential.

\section{Conclusions}

In the first part we discussed the pure Coulomb contribution in heavy
ion collisions using the Glauber approach. Using so called "universal
plots", it is possible, to characterize this Coulomb contributions to
peripheral heavy ion collisions for different experiments. The relevant
parameters are the interaction strength $\eta$, the adiabaticity
parameter $\xi$, and $\theta/\theta_{gr}$. It turns out, that for many
experiments a semi-classical approach to the equivalent photon method
is already a sufficiently good description of the data.

In the second part, we could include the nuclear excitation in a
straight forward way. We used the folding formalism and the optical
potential model, to determine the nuclear transition potential. The
latter result was compared to the DWBA formalism. The introduction of
the simplified "ramp" model makes it possible to clarify the role of
the parameters of the used optical potentials. 

A comparison of the sharp cut--off model results with results obtained
from different parameterisations of the optical potential shows, that
the nuclear contribution can be described within the discussed frameworks.
Therefore the introduced model for the nuclear contribution makes it
possible to estimate the nuclear effects on the electromagnetic
Excitation and Dissociation of nuclei.

\vskip0.5cm 
{\Large \bf Acknowledgment}
\vskip 0.5cm \noindent
We want to thank Tohru Motobayashi and Stefan Typel for
interesting discussions.
 

\newpage
\pagestyle{plain}
\baselineskip 16pt
\vskip 48pt
\doublespace

 
\vskip 2cm
\begin{center}
{\Large \sl \bf Figure Captions}
\end{center}

\begin{itemize}

\item[{\bf Figure 1}:]
$\vert 2 \eta \frac{\Omega_0}{R^2}\vert ^2$ for
$\xi = 0.5$, $\eta = 20$ (dashed) and $\eta=2$ (dotted) upper picture 
and for $\xi = 0.1$, $\eta=3$ (dashed) and
$\eta=1.2$ (dotted) lower one. The solid line shows the result for the semi classical
expression (eq. \protect\ref{omegasc}).

\item[{\bf Figure 2}:]
Same as fig. (1) but for $m = 2$

\item[{\bf Figure 3}:]
Some experiments characterized in a $\xi$--$\eta$--plot.

\item[{\bf Figure 4}:]
Cross section for the Coulomb dissociation of 
$51.9 MeV/nucleon$ $^8B$ beam on a $^{208}Pb$ target. The solid line 
shows the total cross section, while the dashed gives the $E2_{C+N}$, 
and the lower line only the $E2_N$ contribution

\item[{\bf Figure 5}:]
$|\Gamma^{E10}_N(b)|^2$ (left hand side) and 
$|e^{i\chi_N(b)}|^2$ (right hand side) as a function of $\frac{b}{R}$. 
The solid and the dashed line belong to the Woods--Saxon model and 
the simplified ramp model respectively.

\item[{\bf Figure 6}:]
$|f^{E22}_N(\theta)|^2$ as a function of
$\theta /\theta _{gr}$ for different values of $\delta$ (solid line
$\delta=0.5$, dashed line $\delta=0.1$ and dotted line
$\delta=0.02$)

\item[{\bf Figure 7}:]
$|f^{E22}_N(\theta)|^2$ as a function of
$\theta /\theta _{gr}$ for different values of $D$ (solid 
line $D=0.5$, dashed line $D=0.3$ and dotted line $D=0$).

\item[{\bf Figure 8}:]
E2 nuclear cross section for the various models
with the same experiment as in fig. 4. The solid line shows the result
according to eq. (\ref{gammadv}) for a Woods--Saxon Potential, the
dashed line for the ramp model and the dotted line for the sharp
cut--off model (eq. \ref{nsharp}). The parameters used for each
potential are given in the text.

\end{itemize} 

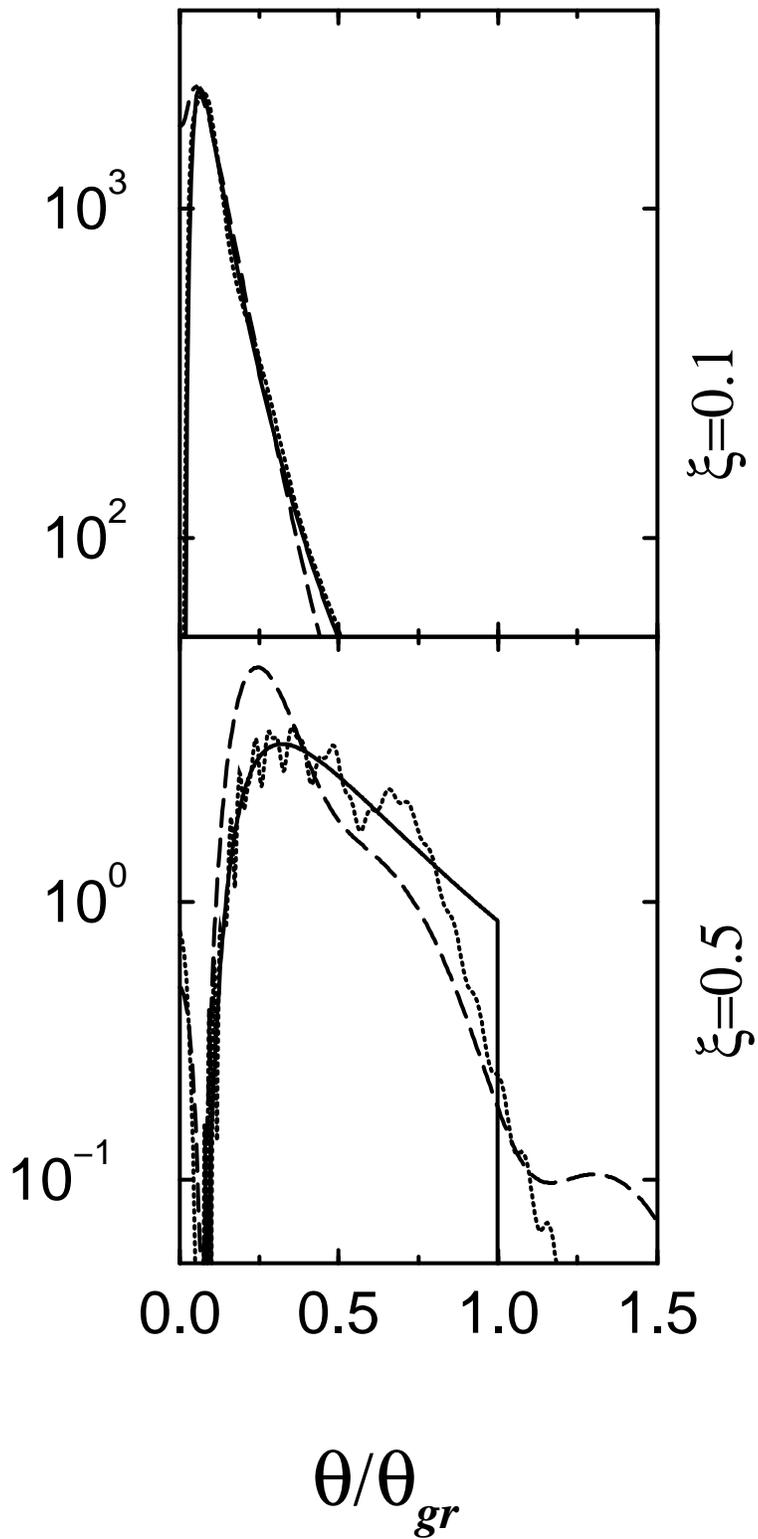
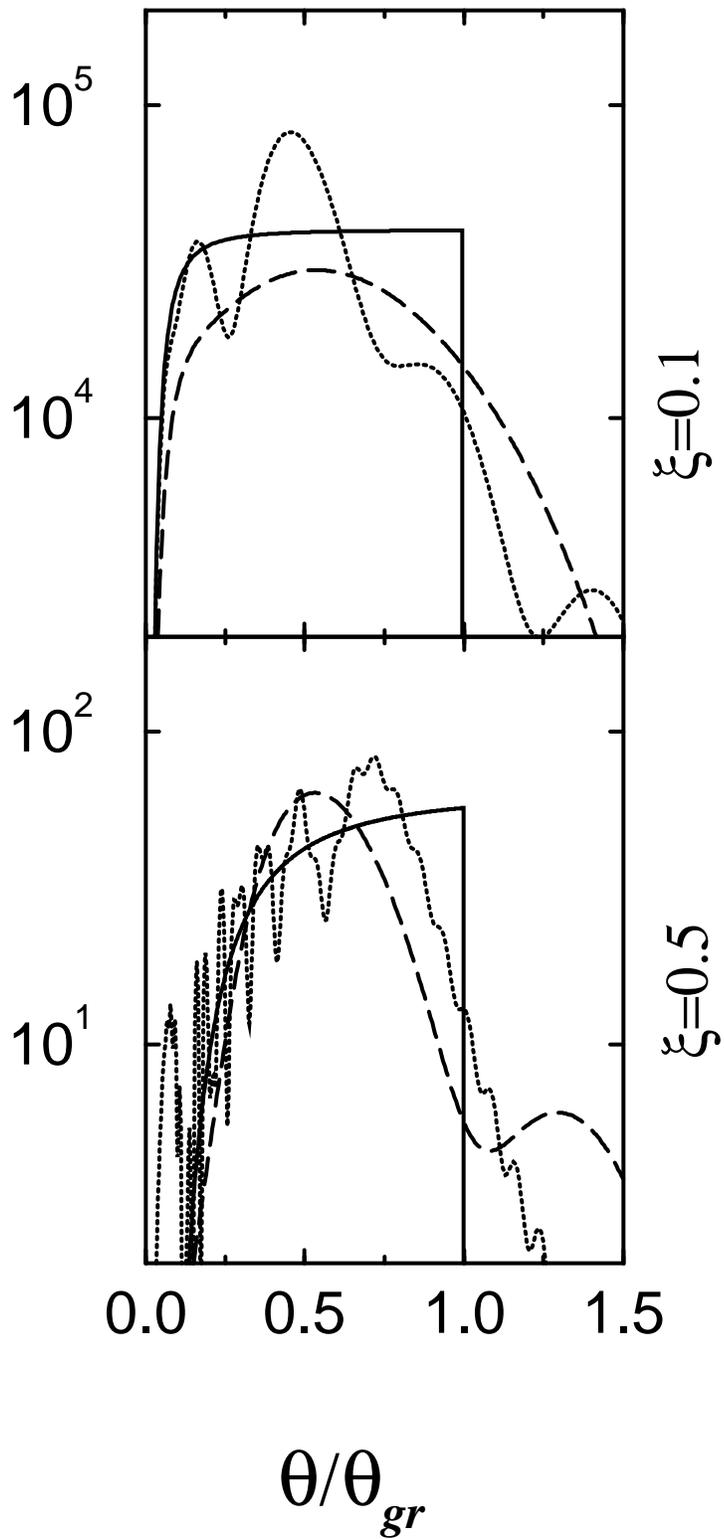
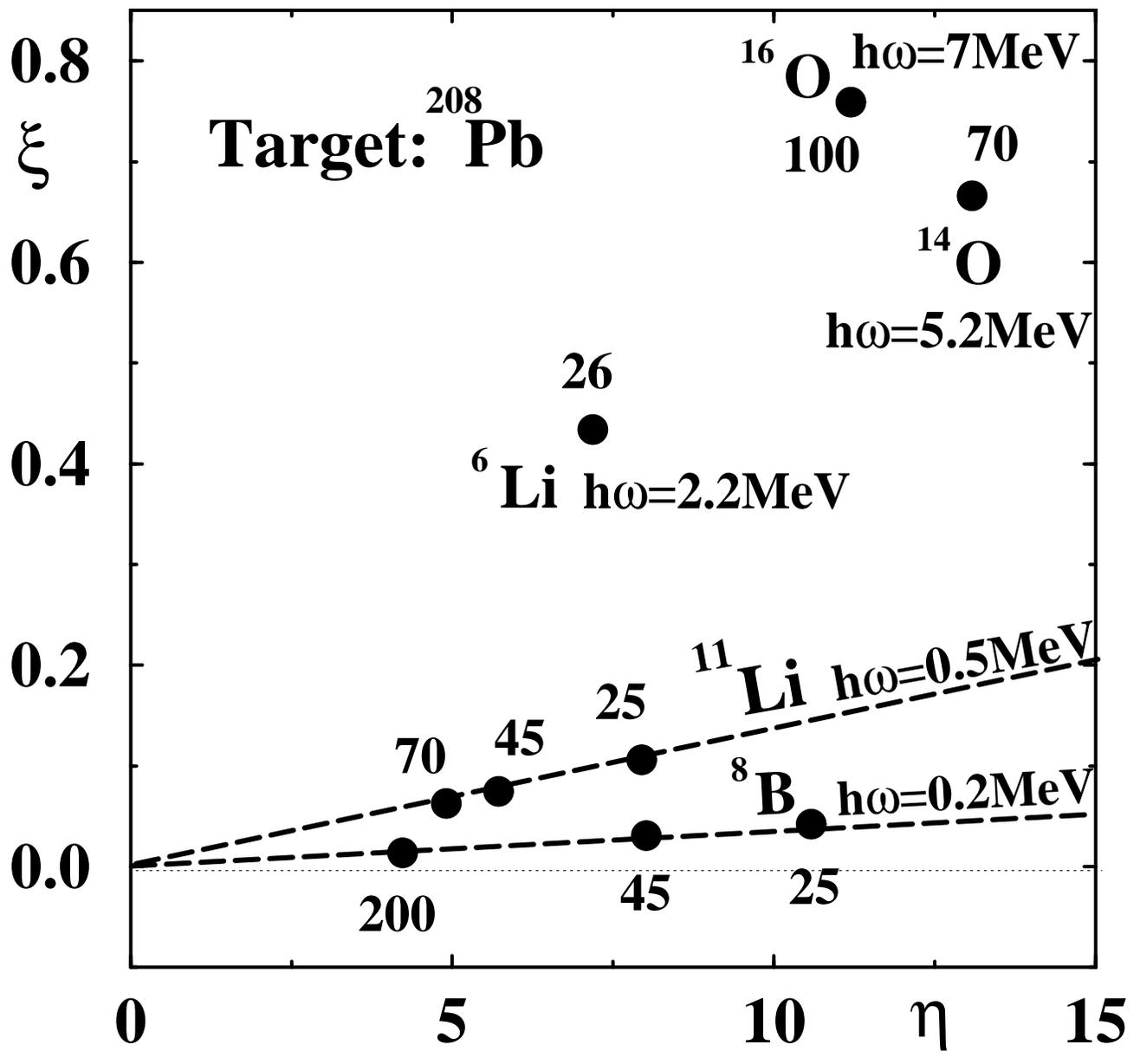
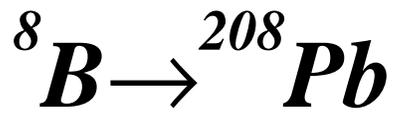
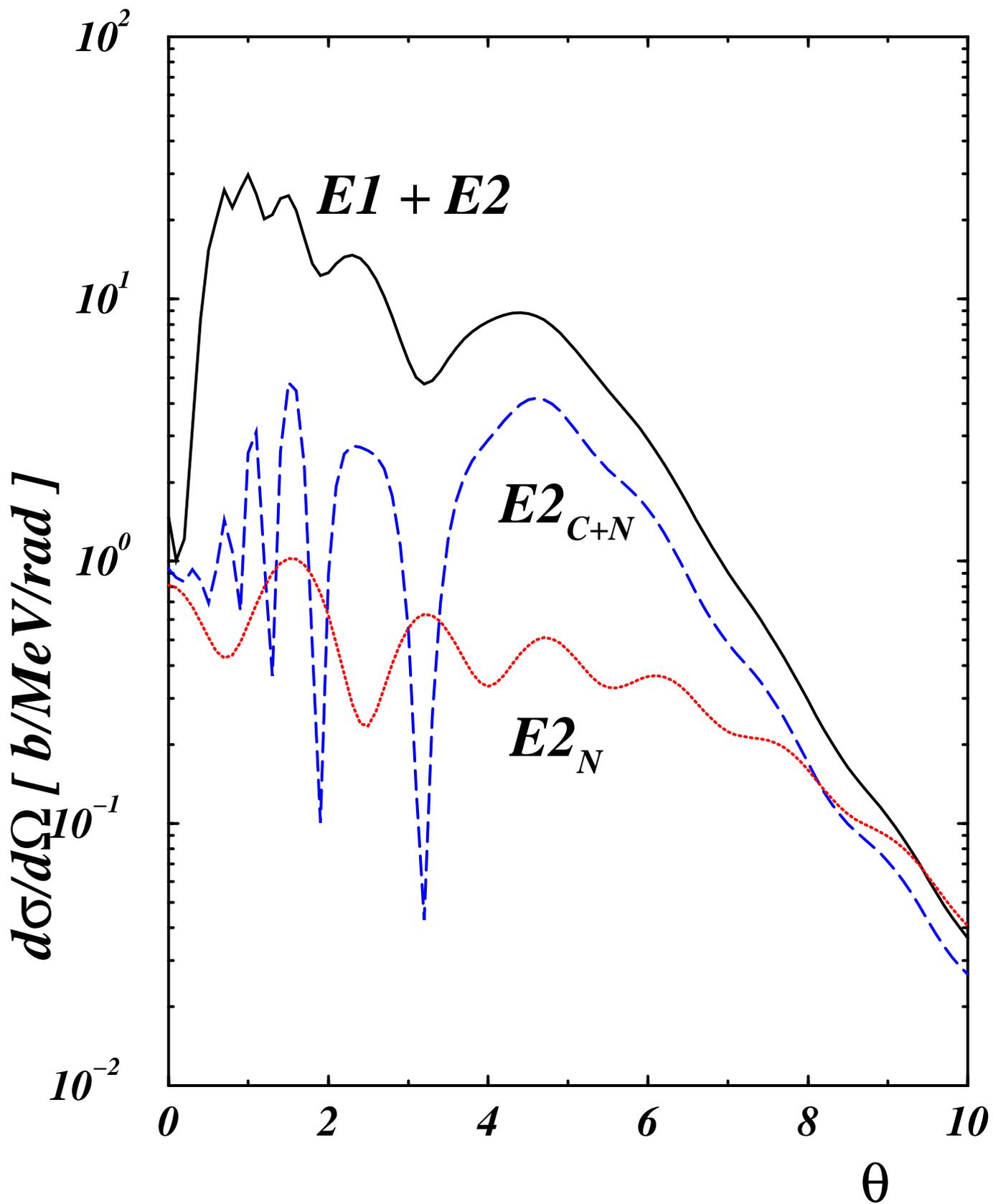
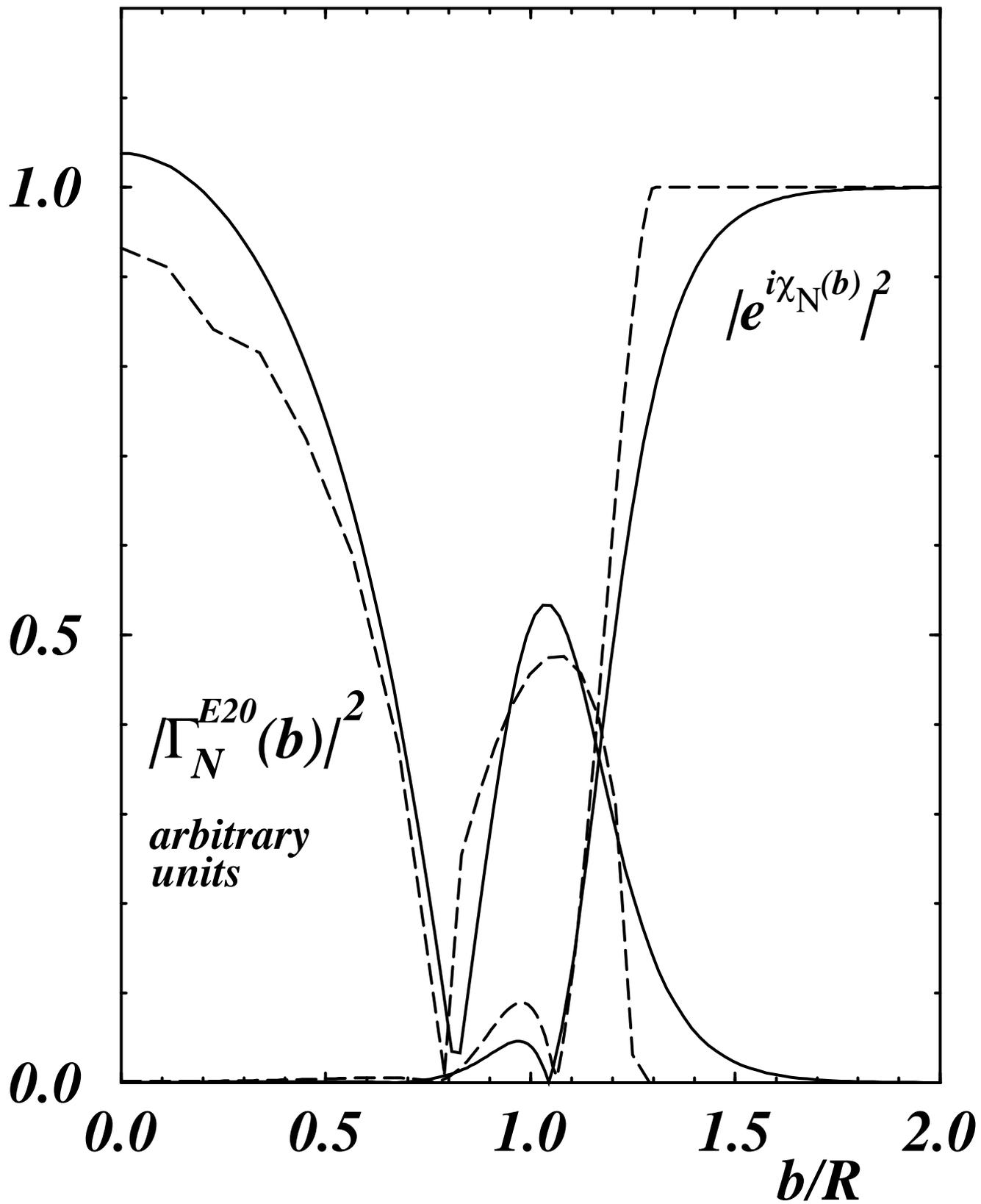
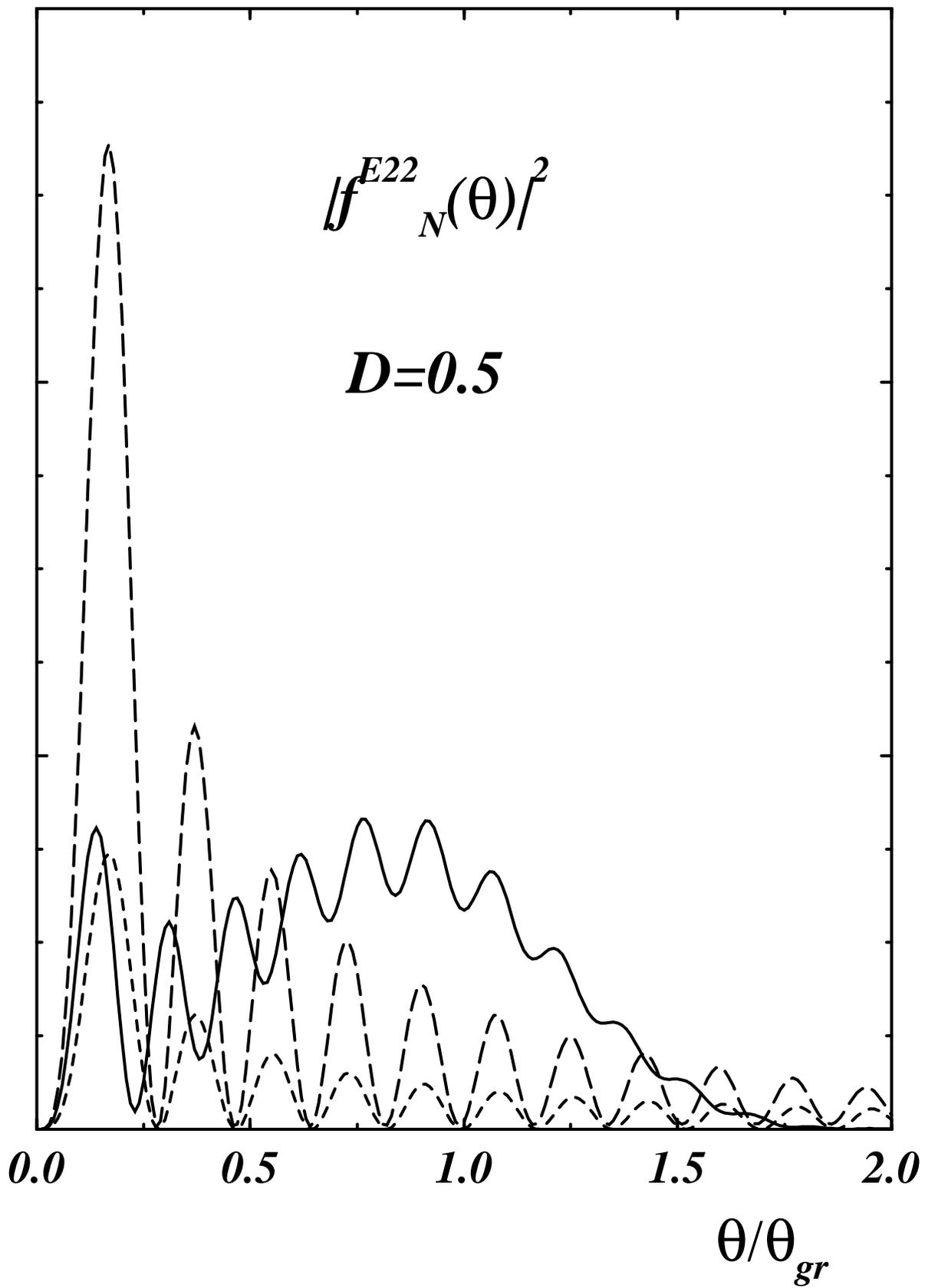
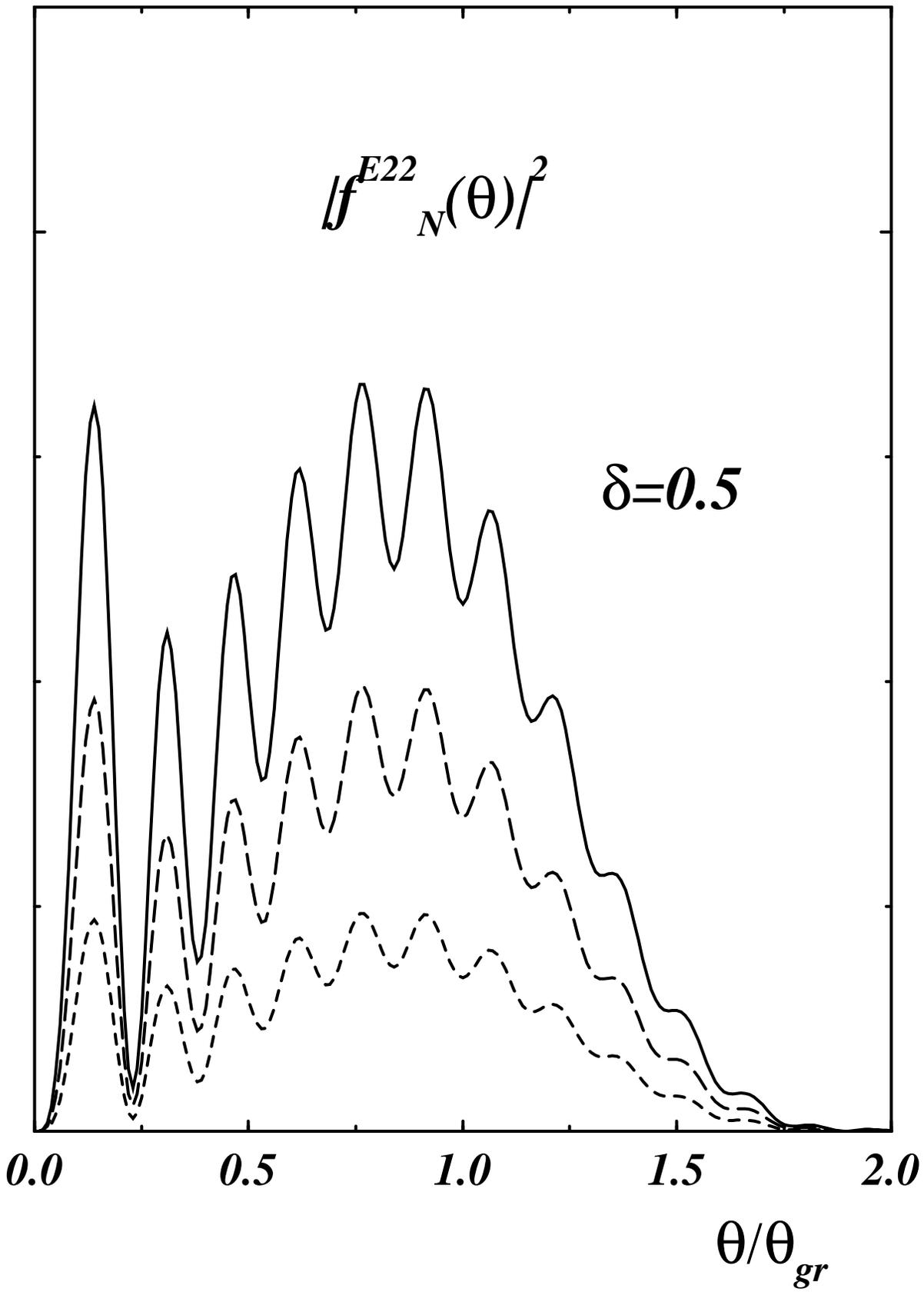
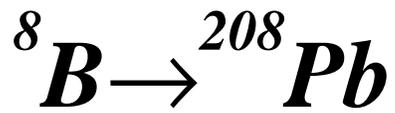
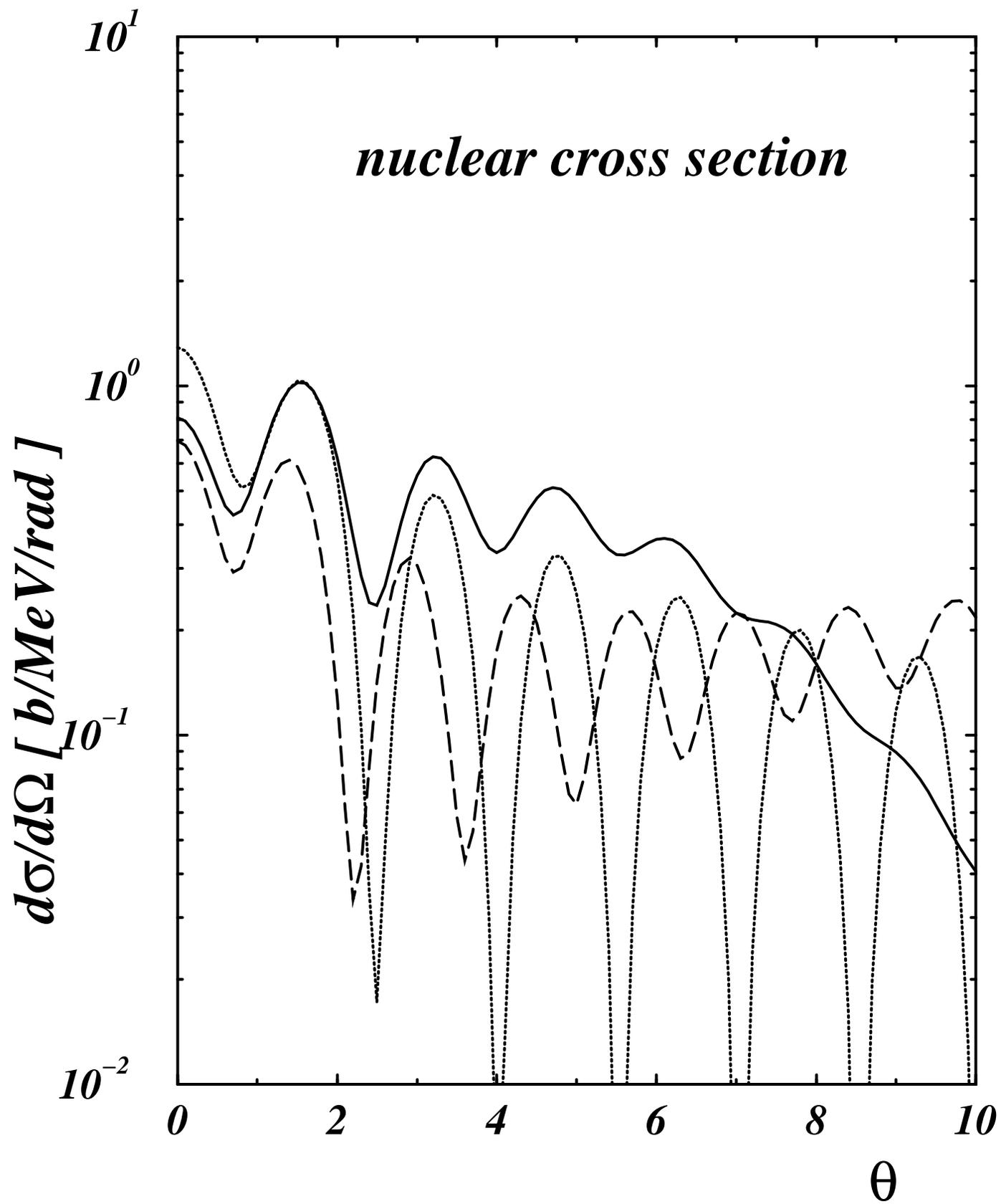
\end{document}